\documentclass[fleqn,usenatbib]{mnras}

\usepackage{amsmath}
\usepackage{color}
\usepackage{xcolor}
\usepackage{graphicx}
\usepackage{grffile}
\usepackage{hyperref}
\usepackage{txfonts}
\usepackage{natbib}
\usepackage{bm}
\usepackage{threeparttable}

\title[Multiwavelength monitoring of NGC~2617]{Long-term multiwavelength monitoring and reverberation mapping of NGC~2617 during a changing-look event}

\author[V. L. Oknyansky et al.]{V. L. Oknyansky$^{1,2,} $\thanks{E-mail: victoroknyansky@gmail.com},
M. S. Brotherton$^{3}$,
S. S. Tsygankov$^{4}$,
A. V. Dodin$^{1}$,
A. M. Tatarnikov$^{1,5}$,
P. Du$^{6}$,
\newauthor
D. -W. Bao$^{6,7}$,
M. A. Burlak$^{1}$,
N. P. Ikonnikova$^{1}$,
V. M. Lipunov$^{1,5}$,
E. S. Gorbovskoy$^{1}$,
V. G. Metlov$^{1}$,
\newauthor
A. A. Belinski$^{1}$,
N. I. Shatsky$^{1}$,
S. G. Zheltouhov$^{1,5}$,
N. A. Maslennikova$^{1,5}$,
J. -M. Wang$^{6,7,8}$,
\newauthor
S. Zhai $^{6,7}$, 
F. -N. Fang$^{6,7}$,
Y. -X. Fu$^{6,7}$,
H. -R. Bai$^{6,7}$,
D. Kasper$^{9}$,
N. A. Huseynov$^{10}$,
J. N. McLane$^{3}$,
\newauthor
J. Maithil$^{3}$,
T. E. Zastrocky $^{3}$,
K. A. Olson$^{3}$,
X. Chen$^{11}$,
D. Chelouche$^{2,12}$,
R.S. Oknyansky$^{13}$,
\newauthor
D. A. H. Buckley$^{14}$,
N. V. Tyurina$^{1}$,
A. S. Kuznetsov$^{1}$,
R. L. Rebolo$^{15}$,
B. -X. Zhao$^{16}$
\\
$^{1}$Sternberg Astronomical Institute, M.V. Lomonosov Moscow State University,  119234, Moscow, Universitetsky pr-t, 13, Russia\\
$^{2}$Department of Physics, Faculty of Natural Sciences, University of Haifa, Haifa 3498838, Israel\\
$^{3}$Department of Physics and Astronomy, University of Wyoming, Laramie, WY 82071, USA\\
$^{4}$Department of Physics and Astronomy, FI-20014 University of Turku, Turku,20014, Finland\\
$^{5}$Faculty of Physics, Moscow M.V. Lomonosov State University, Leninskie gory 1, Moscow, 119991, Russia\\
$^{6}$Key Laboratory for Particle Astrophysics, Institute of High Energy Physics, Chinese Academy of Sciences, 19B Yuquan Road, Beijing 100049, China\\
$^{7}$School of Astronomy and Space Science, University of Chinese Academy of Sciences, 19A Yuquan Road, Beijing 100049, China\\
$^{8}$National Astronomical Observatories of China, Chinese Academy of Sciences, A20 Datun Road, Beijing 100012, China\\
$^{9}$ Department of Astronomy \& Astrophysics, University of Chicago, 5640 South Ellis Avenue, Chicago, IL 60637, USA\\
$^{10}$Shamakhy Astrophysical Observatory, Ministry of Science and Education of Azerbaijan,  Shamakhy, AZ5618, Azerbaijan \\
$^{11}$School of Space Science and Physics, Shandong University, Weihai, Shandong, 264209, China\\
$^{12}$Haifa Research Center for Theoretical Physics and Astrophysics, University of Haifa, Haifa, 3498838, Israel\\
$^{13}$ Israel Space Weather and Cosmic Ray Center, Tel Aviv University, Israel Space Agency, \& Golan
Research Institute, Tel Aviv, 69978,  Israel\\
$^{14}$The South African Astronomical Observatory, P.O. Box 9, Observatory, 7935, South Africa\\
$^{15}$The Instituto de Astrofisica of Canarias Via Lactea, Tenerife, E38205, Spain\\
$^{16}$Shanghai Astronomical Observatory, Chinese Academy of Sciences, Shanghai 200030 China
}

\date{Accepted XXX. Received YYY; in original form ZZZ}

\pubyear{2023}

\begin{document}
\date{Received ... Accepted ...}
\pagerange{\pageref{firstpage}--\pageref{lastpage}}
\maketitle{}

\label{firstpage}

\begin{abstract}
We present the results of photometric and spectroscopic monitoring campaigns of the changing look AGN NGC~2617 carried out from 2016 until 2022 and covering the wavelength range from the X-ray to the near-IR. The facilities included the telescopes of the SAI MSU, MASTER Global Robotic Net, the 2.3-m WIRO telescope, {\it Swift}, and others. We found significant variability at all wavelengths and, specifically, in the intensities and profiles of the broad Balmer lines. We measured time delays of $\sim$ 6 days ($\sim$ 8 days) in the responses of the H${\beta}$  (H${\alpha}$)  line to continuum variations. We found the X-ray variations to correlate well with the UV and optical (with a small time delay of a few days for longer wavelengths). The $K$-band lagged the $B$ band by  14 $\pm$ 4 days during the last 3 seasons, which is significantly shorter than the delays reported previously by the 2016 and 2017--2019 campaigns. Near-IR variability arises from two different emission regions: the outer part of the accretion disc and a more distant dust component. The $HK$-band variability is governed primarily by dust. The Balmer decrement of the broad-line components is inversely correlated with the UV flux. The change of the object's type, from Sy1 to Sy1.8,  was recorded over a period of $\sim$ 8 years.  We interpret these changes as a combination of two factors: changes in the accretion rate and dust recovery along the line of sight.

\end{abstract}




\section{Introduction}

NGC~2617 was first mentioned by Edouard Jean-Marie Stephan in 1885. It is a face-on Sc galaxy at a redshift of z = 0.0142  classified as Seyfert 1.8  (Sy 1.8) \citep{moran1996, Paturel2003} and located in Hydra.  \cite{Shappee13, Shappee14} detected a dramatic change of the object's spectral type that followed a significant brightening of the nucleus discovered by the All-Sky Automated Survey for SuperNovae (ASAS-SN) in April 2013. Prior to 2013, the object had been studied very rarely. Only two spectra were published: one using a slit spectrograph in 1994 \citep{moran1996} and the other  with an objective prism in 2003 (6dFGS).  These old  spectra demonstrated the lack of the broad component in H$\beta$ and very shallow broad H$\alpha$ wings which resulted in a Sy 1.8 classification \citep{Osterbrock1987}. Comparison  with more recent spectral observations  allowed  \cite{Shappee14} to determine a change of spectral type (CL) from Sy 1.8 to Sy 1 between 2003 and 2013. The CL transition can be numerically indicated as -2, that is two classes from 1.8 through 1.5 to type 1, in the ``magnitude'' system introduced by \citet{Runco2016}. The following  IR, optical, UV photometry  and X-ray observations revealed a strong outburst in all wavelengths. Specifically, the X-ray flux in 2013 was significantly brighter than measured by ROSAT \citep{Boller1992}.  This triggered  more intensive investigations  of NGC~2617 from X-ray to radio wavelengths \citep{Mathur2013, Shappee2013b,  Tsygankov2013, Yang2013, Oknyansky16a, Oknyansky16b, Oknyansky16c, Oknyansky17, Oknyansky17b, Oknyansky2018,Giustini2017, Sheng2017, Fausnaugh2017, Fausnaugh2018, Sheng2017, Olson2020, Yang2021, Feng2021}. 

 We began spectroscopic and photometric (IR $JHK$ and optical $BVRI$) monitoring of NGC~2617 in January 2016 to trace potential state changes (see  \citealt{Oknyansky17b} = Paper I, \citealt{Oknyansky17, Oknyansky2018}). Based on archival photometry (MASTER Global Robotic Net \citep{Kornilov2012,Lipunov2010,Lipunov2019,Lipunov2022}, MASTER hereafter), we  proposed that the Seyfert type change had probably occurred between 2010 October and 2012 February. In 2016, while in the bright state, the object revealed a new emission component in the red wing of H$\beta$. Reverberation mapping analysis found that UV emission variations lag  the  X-ray  by about 2-3 days as in the findings of \cite{Shappee14}. Uncorrelated  X-ray/UV events were observed, which can be explained via X-ray  absorption  by  Compton-thick clouds moving across the line of sight (see Paper I). This scenario is  in agreement with an independent result for  NGC~2617: the detection of a redshifted, high-velocity and highly ionised absorber with a column  density $N_{H} \sim$ 10$^{23}$ cm$^{-2}$  \citep{Giustini2017}.
 
  IR variations in the {\it K} band were found to be delayed with respect to the optical ones by $\sim$ 22 days, but variations in the  {\it J} band were delayed by only $\sim$ 3 days. The time delay in the $K$ band is inconsistent with  the  value found by \cite{Shappee14} (see Paper I). 
  From May 2016  we also requested {\it Swift} observations of NGC~2617.  In December 2017, the X-ray flux was detected to be the lowest since monitoring began in 1990-1991.  The object brightened  again in July 2018 \citep{Oknyansky2018} and demonstrated Sy1 type spectra at least until January 2019 \citep{Senarath2021}.

The first reverberation mapping for the optical emission lines was done for NGC~2617 by \cite{Fausnaugh2017} and an H$\beta$-to-continuum time delay of 6.4$\pm$ 0.5 days  was found.  The  reverberation mapping for the continuum variations of the object at different wavelengths (optical, UV and X-ray) during 2014 was published by  \cite{Fausnaugh2018}.
The lags were consistent with geometrically thin accretion-disc models, however, the observed lags were larger than predictions based on standard thin-disc theory by a factor of a few.
The X-ray variability of NGC~2617 (among the  number of selected Seyfert 1.8/1.9 objects) was investigated by \cite{Hernandez2017} and its continuum was modelled by a combination of two  power laws. 

The results of a high cadence velocity-resolved reverberation mapping campaign for the interval from 2019 October to 2020 May were recently published by \cite{Feng2021}. The investigation found the time lags for H$\alpha$, H$\beta$, H$\gamma$ decreasing from the former lines to the latter.  For   H$\alpha$, and H$\beta$  the lag of the line cores were found to be longer than those of the relevant wings, and the peak of the velocity-resolved cross-correlations was slightly blueshifted.

The physical nature of the CL  AGN phenomenon is still mysterious, particularly, the origin of the UV and optical variability, as well as its relation to the X-ray emission, which is not yet clearly established. Therefore variability investigations can be very helpful for understanding the central structures and physical processes in AGNs
 \citep [see discussion and references in][] {Runnoe2016,MacLeod2019,Oknyansky17,Oknyansky2019,Oknyansky2021, Ruan2019, Wang2020}. As soon as the first CL events in AGNs were discovered \citep{Khachikian1971,Pronik1972,Lyuty84} two possible mechanisms were suspected to be responsible for the dramatic changes of the spectral type: strong change of the accretion rate and variable obscuration \citep[see e.g.][]{Malkan1998}. It is possible that these two mechanisms are both involved. If the region of an AGN is obscured for the observer by dust clouds it will have close to a Sy2 optical spectrum. If at some moment a strong outburst happens, then the dust will be sublimated in part and so the dense clouds in the Broad Line Region (BLR) will become directly observable.  If the accretion rate decreases after a while and the object stays in the low state for a few years then the dust can recover in some clouds along the line of sight and  the object will not show  broad emission lines due to low UV radiation and the obscuration by dust clouds in the line of sight. This possible situation was discussed in respect to NGC~2617  and can be considered a prediction (see details in Paper~I and in \cite{Oknyansky2022}). 

Here we present our new investigations of the X-ray/UV/optical/IR continuum and optical emission-line variability of NGC 2617 during October 2016 -- January 2022 using data obtained by the {\it Swift}, MASTER, SAI MSU (Russia), WIRO (USA) and Shandong (China) observatories. We also perform reverberation mapping (RM) of the data to investigate lags between the variable emission components at different wavelengths, including those associated with the H$\beta$ and H$\alpha$ broad emission lines. 

\section{Observations, instruments and data reduction}

In our previous study (Paper I), we reported on the results of spectroscopic and photometric monitoring of NGC~2617 during January -- June 2016. In the present study we combine our old data and our new data obtained partly with the same telescopes, but with some differences and additions. 
\subsection{Spectroscopy}
 First of all, we used data from intensive spectral monitoring with the 2.3-m WIRO telescope (281 epochs in 2016--2022) applying the same methods and procedures as in previous studies (see details in \cite{Du2018}). Some additional spectral data (23 epochs during 2020--2022) were also obtained with 2.5-m telescope of the Caucasus Mountain Observatory (CMO) of SAI MSU (for details about the instrument and methods see \cite{Oknyansky2021}). The WIRO and the CMO spectral data have some systematic differences which are mostly due to different apertures. The CMO data for the $H\beta$ flux were reduced to the WIRO system using linear regression. Spectral data for the H${\beta}$ flux are presented in Table~\ref{Tab4} for the WIRO and CMO data. 
\subsection{{\it Swift} X-ray, UV and optical data}
We present here new {\it Swift} data obtained from June 2016 until January 2022 (120 epochs ) which include X-ray fluxes and UV/optical photometry. The data (through 2017)  were published in part in the conference proceedings by \cite{Oknyansky17b}. The methods were the same as in Paper I, but we used updated calibrations and software and all data were reprocessed uniformly.  All optical and UV {\it Swift} photometry corresponds to a 10 arcsec aperture.
\subsection{Ground-based optical photometry}
We present here the result of the optical photometry obtained with the 0.6-m telescope of the Dibai E.A. Astronomical Station of SAI MSU (DAS)  during 2016--2018 (98 epochs). During September 2019 -- February 2022 we performed intensive (230 dates) $B$ photometry with the new 0.6-m telescope installed at CMO (see details in \cite{Berdnikov2020}). During 2017--2018 the  $B$ photometry (25 epochs) was done the with the 1-m telescope at the Weihai Observatory (WHO) of the Shandong University \citep{Hu2014}. Neither photometric nor spectral data obtained at the Shamakhy Astrophysical Observatory (ShAO) later than May 2016 were included in this study. 
The details  about the methods and instruments used for the optical photometry can be found in  \cite{Oknyansky17,Oknyansky2021} and Paper I.

Unfiltered MASTER photometric data (822 epochs) were obtained at the SAAO and Canarian  MASTER observatories (MASTER-SAAO, MASTER-IAC) during 2015-2022. These data were also  reduced to the {\it B Swift} system (see Table \ref{Tab1}). 

All optical photometry corresponds to a  10 arcsec aperture. All photometric data in $B$ were reduced to the {\it B Swift} system (see Table \ref{Tab1}).  

\subsection{Near-IR photometry} The IR $JHK$ data (194 epochs  during 2016--2022) were obtained with the ASTRONIRCAM camera of the 2.5-m telescope of CMO  \citep{Nadjip2017} and correspond to a 5 arcsec aperture (see Table \ref{Tab3}).

	\begin{figure*}
	\includegraphics[scale=0.9,angle=0,trim=0 0 0 0]{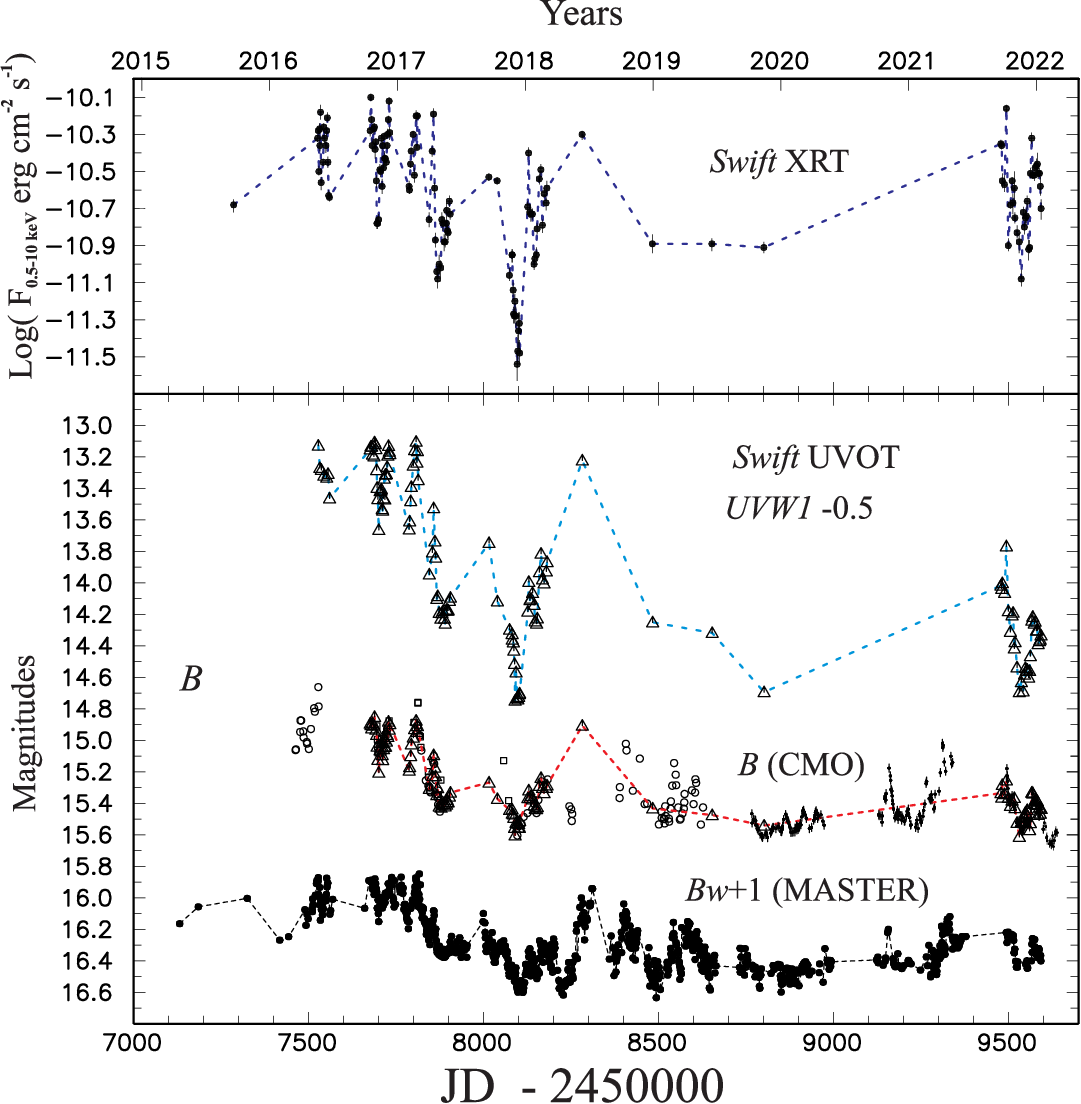}
    \caption{Multi-wavelength observations of NGC~2617 shown for 2016-2021.  {\it Top Panel:} The {\it Swift}/XRT 0.5--10 keV  X-ray flux (in erg cm$^{-2}$ s$^{-1}$). 
 {\it Bottom Panel:} The $Swift$ data in $UVW1$ (connected by blue dashed line) and $B$ (connected by red dashed line) for a 10 arcsec aperture are shown as triangles. Besides the $Swift$ data, the middle curve represents the $B$ data obtained with different telescopes: the 0.6-m telescope of the DAS (open circles), the 1-m telescope of the WHO (boxes), the RC600 of the CMO (nightly mean estimates, dots with error bars). All the $B$ data were measured in an aperture of 10 arcsec and reduced to the $Swift$ $B$ system. The nightly mean MASTER unfiltered data (for 2015--2022) reduced to the Swift $B$ system are shown in the bottom curve (small filled circles).
   }
    \label{fig1}
\end{figure*}

	\begin{figure}
	\includegraphics[width=\columnwidth]{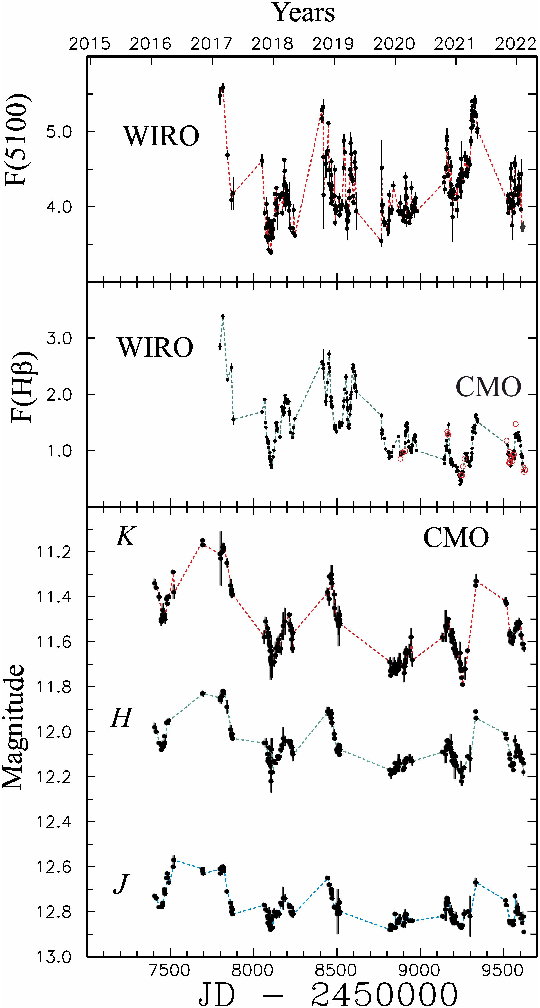}
    \caption{Multi-wavelength and H${\beta}$ variations of NGC~2617 shown for 2016-2022 only.  {\it Top Panel:} The 5100\AA~ continuum variations from the WIRO spectral observations in units 10$^{-14}$erg s$^{-1}$cm$^{-2}$\AA$^{-1}$ (red dashed line).   {\it Middle Panel:} The broad H${\beta}$ emission variations from the WIRO (points connected by green dashed line) and the CMO (red circles) spectral data.  The light curve is given in fluxes (in units 10$^{-13}$erg s$^{-1}$cm$^{-2}$ ). 
 {\it Bottom Panel:}  The IR light curves in $J$ (blue dashed line), $H$ (green dashed line) and $K$ (red dashed line)  obtained with the 2.5-m telescope CMO (for an aperture of 5 arcsec).
   }
    \label{fig2}
\end{figure}


\begin{table}
\centering
\caption{The WIRO and CMO  H${\beta}$ fluxes (in units 10$^{-13}$erg s$^{-1}$cm$^{-2}$. ) (Note: The full version of this table is available in its entirety in machine-readable form.)}

\begin{tabular}{cccc} \hline
   J.D.-2450000& Flux in H${\beta}$& observatory  \\
   \hline
7797.72 &  2.849  $\pm$ 0.063    & WIRO\\
7815.67 &  3.385  $\pm$ 0.043    & WIRO\\
7842.68 &  2.259  $\pm$ 0.032    & WIRO\\
7866.62 &  2.478  $\pm$ 0.079    & WIRO\\
7877.65 &  1.550  $\pm$ 0.086    & WIRO\\
...&  ... & ...\\ \hline
\end{tabular}
 \label{Tab4}
\end{table}

\begin{table}
\centering
\caption{$B$ photometry obtained with the 1-m telescope (WHO), the 0.6-m telescope (DAS), the RC600 telescope (CMO), MASTER (MAS) and the $Swift$ $B$ photometry (Swi). All data are reduced to the {\it Swift} $B$ system. 
(Note: The full version of this table is available in its entirety in machine-readable form.)}

\begin{tabular}{cccc} \hline
   J.D.-2450000& mag & comment  \\
   \hline
7132.00 & 15.164  $\pm$ 0.030 &    MAS\\ 
7185.00 & 15.057  $\pm$ 0.030 &    MAS\\
7325.00 & 15.003  $\pm$ 0.030 &    MAS\\
7418.00 & 15.269  $\pm$ 0.030 &    MAS\\
7443.00 & 15.247  $\pm$ 0.030 &    MAS\\
...&  ...\\ \hline
\end{tabular}
 \label{Tab1}
\end{table}

\begin{table}
\centering
\caption{$JHK$ photometry obtained with the 2.5-m telescope (CMO). (Note: The full version of this table is available in its entirety in machine-readable form.)}

\begin{tabular}{cccc} \hline
   J.D.-2450000& mag & band  \\
   \hline
7409.46 & 11.340  $\pm$ 0.020 &    K\\
7409.46 & 11.980  $\pm$ 0.020 &    H\\
7409.46 & 12.730  $\pm$ 0.010 &    J\\
7418.33 & 11.360  $\pm$ 0.010 &    K\\
7418.33 & 12.000  $\pm$ 0.010 &    H\\
...&  ... & ...\\ \hline
\end{tabular}
 \label{Tab3}
\end{table}

\section{Light curves and time-series analysis}
\subsection{Light curves}

Light curves for X-ray, UV, optical, IR, H${\beta}$ and continuum at $\lambda$5100 (WIRO)  variations are given in Fig.~\ref{fig1} and Fig.~\ref{fig2}.
The variations before July 2018 were discussed mostly in the Introduction and our previous publications. The variations at different wavelengths (also IR and H${\beta}$) are well correlated, at least for the events occurring on relatively short time scales (weeks and months). Our light curves are also in a good agreement with the photometry for the interval from 2019 October to 2020 May published by \cite{Feng2021}. As can be seen from the X-ray, UV and optical light curves, since the second half of 2018 until the start of 2021 the brightness in all wavelengths was mostly low with low-amplitude (0.2--0.4 mag)  fast and short (on a time scale of weeks) brightenings (here and below estimates of the amplitude are given for $B$ band). From the spring through October 2021 the object had several relatively strong brightenings (with amplitude $\sim$~0.5--0.7 mag on a time scale of a month). For the last season 2021-2022 the mean optical  brightness  decreased by $\sim$~0.2 mag with some fluctuations (with amplitude $\sim$~0.2 mag on a time scale of months). This change in variability stimulated us to request new $Swift$ observations. The new set of the {\it Swift} monitoring (September 2021 -- January 2022) revealed  relatively high amplitude of variability in all wavelengths. The maximal levels of the X-ray fluxes were close to that for the high state seen in light curves during the first years but in the UV and optical bands the brightness was mostly below the mean. If we look at these light curves, we can clearly see the difference between the X-ray and UV/optical variations: the X-ray flux has nearly constant mean level along the whole light curve whereas some decreasing trend is present in the longer wavelengths from the UV to optical bands. Due to this difference, the UV/X-ray flux ratio must have also had a strong decrease in 2016--2022.

\subsection{Reverberation mapping method}
In order to investigate possible lags between continuum variations in different wavelengths (from X-ray, UV to near-IR), as well as between continuum and broad H${\beta}$, we used the  MCCF method \citep{Oknyansky1993}, which is an upgraded version of the ICCF (traditional interpolation cross-correlation techniques 
\citep{Gaskell1986,Gaskell1987, Gaskell1988,Peterson1998,Peterson2004}) for two unevenly spaced data sets. The methodology of our analysis is the same as described in our previous papers (see, e.g., Paper I and \cite{Oknyansky2014a, Oknyansky2014b, Oknyansky2018, Oknyansky2021}). There is a publicly available Python code  \citep{Sun2018} of the ICCF (PyCCF) adopted from the code written by B. Peterson \citep{Peterson1998}. The PyCCF slightly differs from the original implementation of the method \citep{Gaskell1987, Gaskell1988}. In the primary version, in order to perform the cross-correlations one has to interpolate linearly between  measurements and then extend the measurements backward and forward in time at a constant level equal to the first and last values. So, the numbers of the pairs for correlation at each lag are the same. Such interpolations are performed firstly for the first data set $a(t)$ while the  data in the second data set $b(t-\tau)$ (where $\tau$ is a time delay) are taken as they are. Then the same interpolating procedure is applied to the second data set $b(t)$ while the data in $a(t+\tau)$ are used without interpolation. So, the $CCF(\tau)$ is calculated twice for each time delay $\tau$ and then the final result is taken as an average of the two. For the PyCCF code, the extrapolation is not used to calculate the ICCF function and, so, the number of pairs of values used for correlation at each time shift is not constant. In our MCCF method we also have different numbers of pairs for each delay and moreover we remove from consideration some interpolated points in $a(t)$ which are separated in time by more than some limit value (MAX) from the nearest observation points in the  $b(t-\tau)$   data set. The MAX is a trade-off value which is chosen taking into account the data sampling, the variability time scales, and the need to have enough data to be able to perform RM. In the MCCF we first of all strive to reduce the contribution made by interpolation errors as well as to introduce a minimum number of free parameters. We performed test calculations with our code  with real  observational data to see how robust the results are depending on the selected values of MAX in intervals between 2 -- 20 days and did not find any variations bigger than the error values. The bigger MAX value tend to give more  noise from the interpolated points in the gaps and that is a reason for lower correlation coefficients. One more difference of our MCCF method from the ICCF and PyCCF codes is that we  interpolate just the data sets that have better accuracy and cadence. If the data sets are nearly the same for these parameters, we get the final $CCF(\tau)$ in a similar way as the ICCF does. Examples of using this method compared to other RM methods (JAVELIN, ICCF) can be found in a number of our publications (see, e.g., Paper I and \cite{Oknyansky2021}).  Since the results obtained by these methods are in good agreement, here we have used only the MCCF to perform RM. Following the analogy of the PyCCF, we have developed a Python code that implements the MCCF method (for the PyMCCF code and comments see \cite{oknyansky2022b}).

\begin{table*}
\centering
\caption{Results of the RM analysis with the MCCF code. The lags  for the $X$ data sets are measured with respect to the Y data and are expressed in days (in rest frame). A positive lag means the Y set leads the X variability.  The correlation values in the peaks $R_{max}$ are given. The confidence limits and centroid cross-correlations are presented. Our preferred $\tau_{peak}$ values are given in bold. All delays are reduced to the rest system.}

\begin{tabular}{ccccccccc} \hline
  X &  Y        & J.D.-2450000  & Date            & MCCF    \\
  &&&& $R_{max}~~~~~~~\tau_{peak}~~ \tau_{cent}$ \\
  \hline \smallskip

I($B$)(CMO+$Swift$) & I(X-ray)&9492--9592& Sep.2021--Jan.2022 &0.864~~~${\bf1.5^{+0.3}_{-0.1}}~~~ 1.8^{+0.4}_{-0.2}$   \\ \smallskip

I($B$)(CMO+$Swift$) & I(UVW1)&9492--9592& Sep.2021--Jan.2022 &0.931~~~${\bf0.0^{+1.1}_{-0.3}}~~~ 0.4^{+0.7}_{-0.6}$   \\ \smallskip

I($B$)($Swift$) & I(UVW1)&7676--8181& Oct.2016--Mar.2018 &0.994~~~${\bf0.2^{+0.3}_{-1.0}}~~~ 0.7^{+0.5}_{-0.5}$   \\ \smallskip

 I($UVW1$)&I(X-ray)& 6414--6756& Mar.2013--Apr.2014 &0.883~~~${\bf2.1^{+1.3}_{-0.3}}~~~ 3.3^{+0.1}_{-0.1}$     \\ \smallskip
 
 I($UVW1$)&I(X-ray)& 7676--8181& Oct.2016--Mar.2018 &0.910~~~${\bf1.5^{+1.8}_{-1.9}}~~~ 3.3^{+0.3}_{-0.4}$     \\ \smallskip
 
I($UVW1$)&I(X-ray) & 9478--9592& Sep.2021--Jan.2022 &0.883~~~${\bf1.6^{+0.2}_{-0.5}}~~~ 1.2^{+0.2}_{-0.2}$     \\ \smallskip

I($J$) (CMO) &  I($B$) (CMO) &8766-9337& Oct.2019--May 2021 & 0.913~~${\bf ~4.8^{+4.6}_{-2.4}}~~ ~8.1^{+1.6}_{-4.7}$\\ \smallskip

I($H$) (CMO) &  I($B$) (CMO) &8766-9337& Oct.2019--May 2021 & 0.898~~${\bf 15.3^{+4.7}_{-1.0}}~~ 18.1^{+1.0}_{-5.9}$ \\ \smallskip

I($K$) (CMO)&  I($B$) (DAS) &7405-7557& Jan.2016--Jun. 2016 & 0.899~~~${\bf23.9^{+5.4}_{-5.1}}~~22.5^{+3.2}_{-2.7}$\\ \smallskip 
 
I($K$) (CMO)&  I($B$) (DAS) &7811-8627& Feb.2017--May 2019 & 0.923~~~${\bf27.5^{+3.5}_{-6.7}}~~21.8^{+7.6}_{-3.5}$\\ \smallskip 

I($K$) (CMO) &  I($B$) (CMO) &8766-9337& Oct.2019--May 2021 & 0.896~~${\bf 14.0^{+5.3}_{-3.4}}~~ 18.5^{+2.0}_{-5.2}$ \\ \smallskip

H${\beta}$ (WIRO) &  I($B$) (MASTER) &7797-9592& Feb.2017--NOV. 2021 & 0.798~~~${\bf9.2^{+0.4}_{-1.6}}~~~ 7.8^{+1.0}_{-0.9}$\\ \smallskip 

H${\beta}$ (WIRO) &  I($B$) (DAS) &7811-8627& Feb.2017--May 2019 & 0.915~~~${\bf6.3^{+1.6}_{-0.7}}~~~ 4.7^{+0.4}_{-0.4}$                     \\ \smallskip

H${\beta}$ (WIRO) &  I($B$) (CMO) &8766-9618& Oct.2019--Feb.2021 & 0.427~~~${\bf6.2^{+1.1}_{-0.4}}~~~ 6.5^{+0.2}_{-0.3}$                     \\ \smallskip

H${\beta}$ (WIRO) &  I($B$) (CMO) &8766-8978& Oct.2019--May 2020 & 0.780~~~${\bf4.3^{+0.8}_{-0.7}}~~~ 4.6^{+1.1}_{-0.7}$                     \\ \smallskip

H${\beta}$ (WIRO) &  I($B$) (CMO) &9143-9342& Oct.2020--May 2021 & 0.917~~~${\bf7.2^{+0.4}_{-1.9}}~~~ 7.3^{+0.8}_{-0.4}$                     \\ \smallskip

H${\beta}$ (WIRO) &  I($B$) (CMO) &9524-9618& Nov.2021--Feb. 2022 & 0.921~~~${\bf5.6^{+0.4}_{-1.8}}~~~ 4.9^{+0.5}_{-0.6}$                     \\ \smallskip
H${\alpha}$ (WIRO) &  I($B$) (DAS) &7811-8627& Feb.2017--May 2019 & 0.800~~~${\bf8.4^{+4.4}_{-0.7}}~~~ 7.0^{+0.5}_{-0.1}$                     \\ \smallskip
H${\alpha}$ (WIRO) &  I($B$) (CMO) &9143-9342& Oct.2020--May. 2021 & 0.707~~~${\bf7.4^{+2.3}_{-2.8}}~~~ 9.6^{+2.3}_{-2.8}$                     \\ \smallskip
H${\alpha}$ (WIRO) &  I($B$) (CMO) &9524-9618& Nov.2021--Feb. 2022 & 0.801~~~${\bf6.6^{+2.3}_{-1.7}}~~~ 7.2^{+2.8}_{-1.3}$                     \\ \smallskip
 
\smallskip    

\end{tabular}
 \label{tab5}
\end{table*}

\subsection{Time delay measurements}
The emission-line time lags multiplied by the speed of light ($c$) provide the emissivity-weighted size scales of the regions corresponding to the variable broad lines. In case of time lags associated with the broadband X-ray/UV/optical continuum responses, they may correspond to an effective radius of the AD (accretion disc), although there may be a component of continuum from the Broad Line Region (BLR) as well \citep[see, e.g.,][]{Korista2001, Lawther2018, Doron2019}.

 The RM for continuum flux variations in the optical ($B$) and $UVW1$ bands confirms the results of visual inspection -- these variations show a very high correlation with a time delay of about zero (see Table~\ref{tab5}). A small time delay of a few hours is possible (see  the values of the centroid estimates) but due to the gaps in the $Swift$ observations it can not be found with high confidence. The result is in agreement with previous results from \cite{Fausnaugh2018}. 
 
\begin{figure}
\includegraphics[width=\columnwidth]{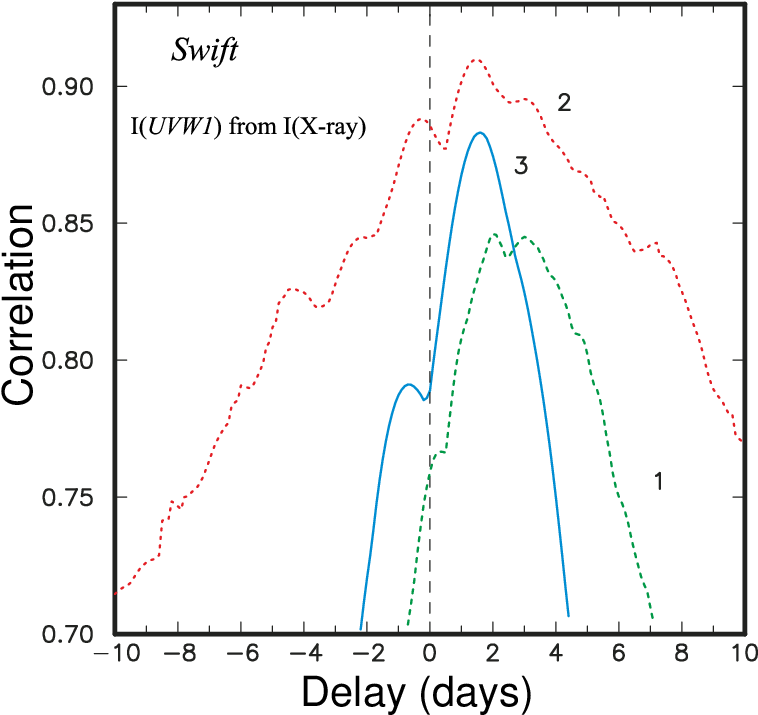}
    \caption{RM by the MCCF method (MAX=5 days) for $UVW1$  relative to X-ray flux variations ($Swift$ data) for 3 independent intervals: 1 (green dashed line) -- old data of 2013--2014, 2 (red dot line)  --  the data of 2016-2018 and  3 (blue solid line) -- the last set of data for September 2021--January 2022.} 
   \label{fig7}
\end{figure}

\begin{figure}
\includegraphics[width=\columnwidth]{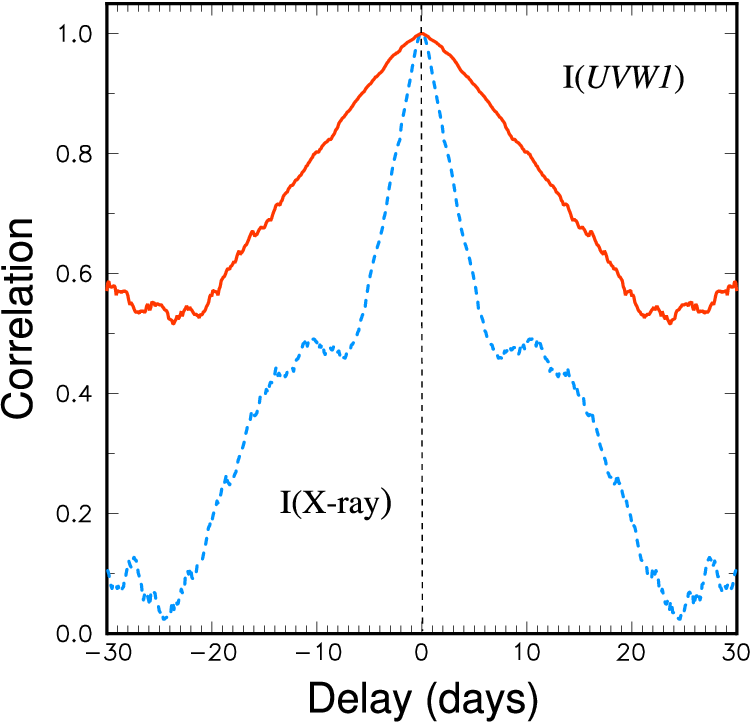}
    \caption{Auto-correlation functions  for I($UVW1$) (red solid line) and I(X-ray) (blue dashed line) for 2016--2022 ($Swift$ data).} 
   \label{fig7-ad}
\end{figure}

 The variations of the X-ray flux led the I($UVW1$) variations by $\sim$ 2.1 days during 2013--2014 (in agreement with what was found by \citealt{Shappee14}), but during 2016--2022 the time delay was $\sim$ 1.5 days in different intervals (see Table~\ref{tab5} and Fig.~\ref{fig7}). This result can not be an artefact connected with broad Auto-Correlation Function (ACF) for I(X-ray) as suspected by \cite{Kammoun2021}. In general, the CCF for input F1(t) and output F2(t) functions is a convolution of the response function $\Psi (\tau)$  for the emission region and ACF(F1(t)) for primary radiation (see Eq.~\ref{ccf} and \citealt{Oknyansky15}).

\begin{equation}
CCF_{F1(t) F2(t)}\sim \int_{-\infty}^{\infty} \Psi (\tau) ACF_{F1(t)}(\tau - t) d\tau,
\label{ccf}
\end{equation}
where $\Psi(\tau)$ is defined by the structure and physical properties of the emitting medium and is the response of the medium to an F1(t) impulse in the form of the $\delta$-function. The ACF for I(X-ray) variations was not too broad for our data (see Fig.~\ref{fig7-ad}) and has no significant local secondary peaks at the $\pm$10 days time delay space. Also the ACF for I($UVW1$) is significantly broader than that for I(X-ray), which is in agreement with the time delay found and with the likely different location and sizes of these two emission regions expected for the case when the X-ray corona radiation is reprocessed by a more extended structure \citep[e.g.,][]{Collier1999}. The fact that the delay of UV variations relative to the X-rays is systematically positive in different time intervals and for a number of AGNs supports the reliability of the delay found in this study.

\begin{figure}
	\includegraphics[width=\columnwidth]{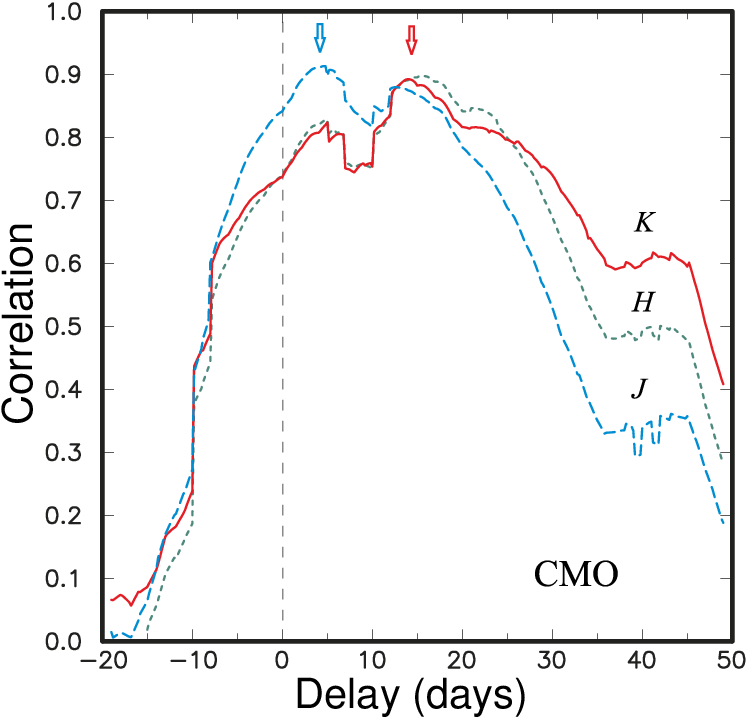}
    \caption{RM by the MCCF method (MAX=5 days) for $J$ (blue dashed line), $H$ (green dashed line), and $K$ (red line) relative to the optical continuum I($B$) (CMO data). The blue and red arrows mark maxima respectively near the delays of about 3 days ($J$) and about 14 days ($HK$).} 
   \label{fig7b}
\end{figure}

\begin{figure}
	\includegraphics[width=\columnwidth]{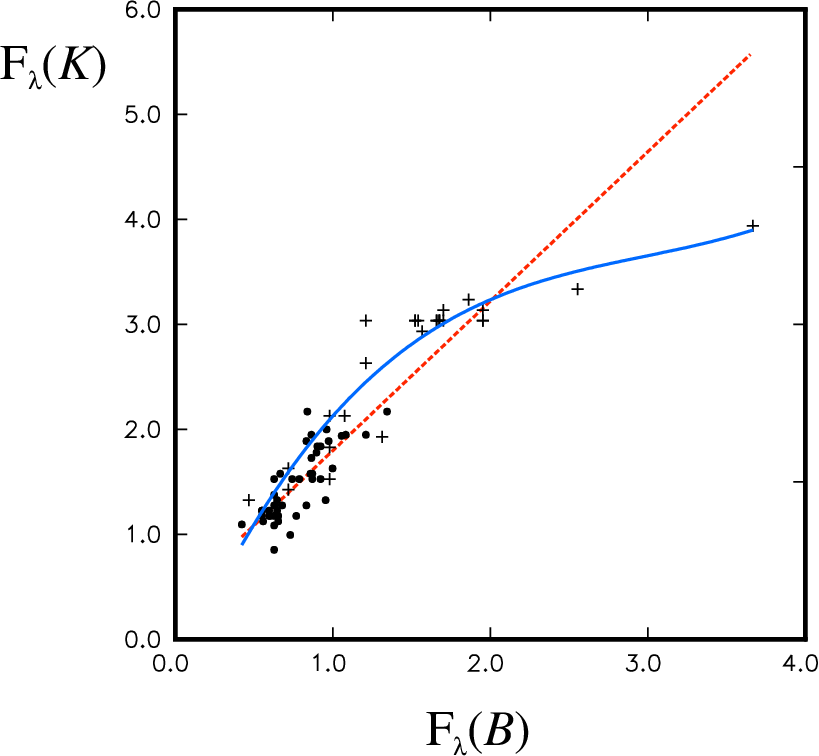}
    \caption{The relation of  $F_{\lambda}(K)$ from $F_{\lambda}(B)$.
Absolute fluxes are in units of $10^{-15}$~erg~cm$^{-2}$ sec$^{-1}$ {\AA}$^{-1}$. The
host galaxy fluxes were subtracted.  The data are for two intervals: filled circles for 2019-2021 when the time delay was 14 days; crosses for seasons 2017-2019 when the time delay was 27.5 days.  The red dashed line is the regression for points with the delay of 14 days 
and  the blue line is polynomial fitting to the data with the time delay 27.5 days.
} 
   \label{ibk}
\end{figure}

The $K$-band lagged the $B$-band by $\sim$14 $\pm$ 4 days during the last 3 seasons, which is significantly smaller than the delays of $\sim$28 $\pm$ 5 days for 2017--2019 (Fig.~\ref{fig7b} and Table~\ref{tab5}) and  $\sim$24 $\pm$ 4 days for 2016 estimated in our previous publication (Paper I).  The relations for F$_{\lambda}(\it{K})$ from F$_{\lambda}(\it{B})$  at these 2 intervals are presented at Fig.~\ref{ibk}. (The time delays, host galaxy contamination were removed, absolute calibration was done using \citep{Tokunaga2005, Straizys1975}). The nonlinear dependence between these fluxes well seen at the plot is also known for some other objects \citep[see e.g.][]{Oknyansky1999b} and may be due to dust sublimation and recovering processes (see more details below in Discussion). 

In the cross-correlation curves for $JHK$ and optical band (Fig.~\ref{fig7b}, there are two local peaks corresponding to different time delays: $\sim$ 6 days and $\sim$ 15 days (Fig.~\ref{fig7b}). The same properties of near-IR emission were found for some other AGNs \citep[e.g., for NGC~4151, see][]{Oknyansky2019b}.

The $JHK$ radiation is believed to possibly have two different origins \citep[see, e.g.,][]{Shnulle2013}: radiation from the outer part of the AD or the inner BLR \citep{Doron2019} which dominates in the $J$-band, and thermal re-emission by dust as  the $HK$ variability is mostly due to the radiation coming from the dust.

\begin{figure}
	\includegraphics[width=\columnwidth]{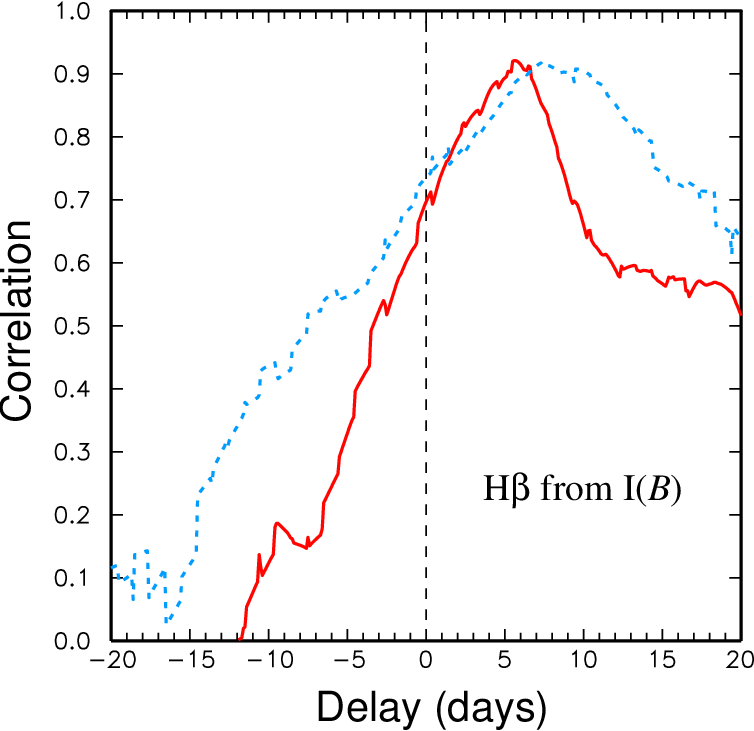}
    \caption{RM by the MCCF method (MAX=2 days) for H$\beta$ (WIRO data) relative to optical continuum I($B$) (CMO data) for two different seasons: 2020--2021 (blue dashed line) and 2021--2022 (red solid line) .
    } 
   \label{fig8}
\end{figure}

The RM results for H${\beta}$ relative to optical continuum I($B$) calculated for different observational seasons is presented in Fig.~\ref{fig8}  (see also Table~\ref{tab5}). The mean value of the time delay was found to be about 6 days, which is in agreement with that found by \cite{Fausnaugh2017} and \cite{Feng2021}. The time delay for H${\alpha}$ is bigger than that for H${\beta}$ by $\sim$ 2 days (see Table~\ref{tab5}) and is in agreement with the delay found by \cite{Feng2021}. The correlations for each season are significantly higher than if the RM is performed for  several seasons together. This can be explained by different regression relations for each season due to variable dust obscuration. The relation can be different also due to asymmetric response function in combination with different trends in each season.

\begin{figure*}
	\includegraphics[scale=0.9,angle=0,trim=0 0 0 0]{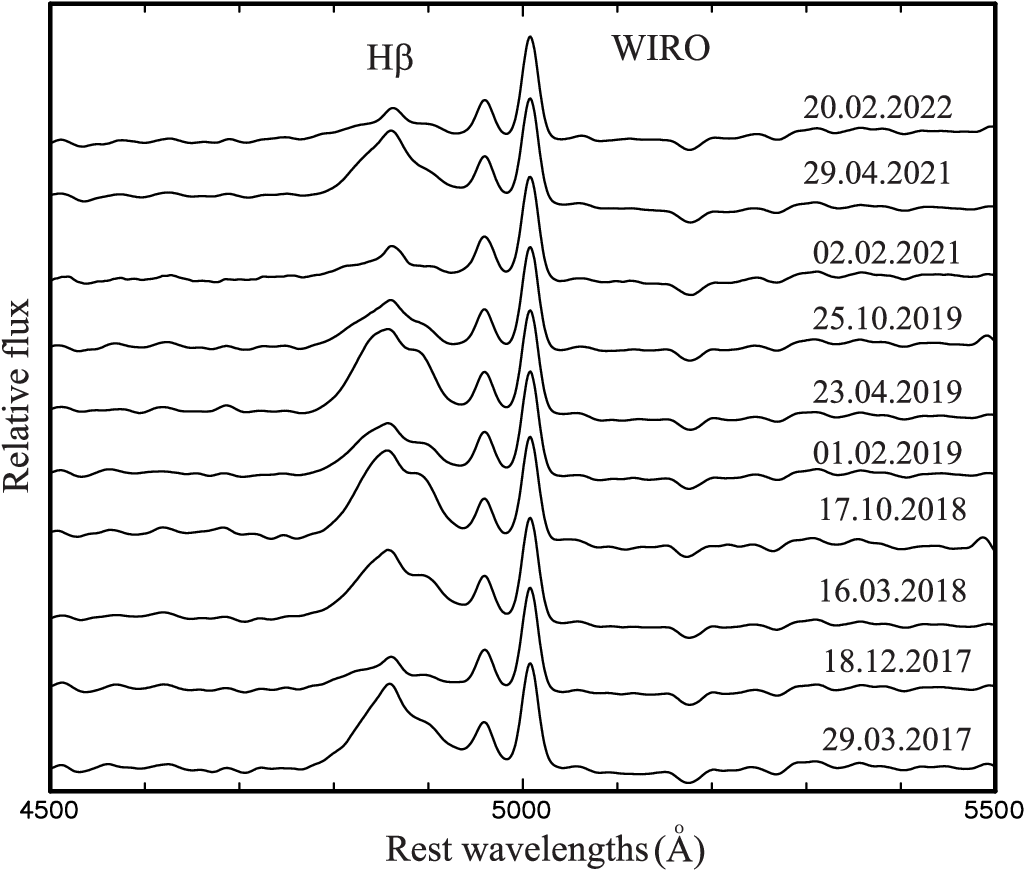}

    \caption{The H${\beta}$ spectral region in the selected spectra obtained at WIRO. The spectra are normalized to the [O III]$\lambda$5007 intensity and shifted vertically for comparison.}
    \label{fig3}
\end{figure*}

\begin{figure*}
	\includegraphics[scale=1.2,angle=0,trim=0 0 0 0]{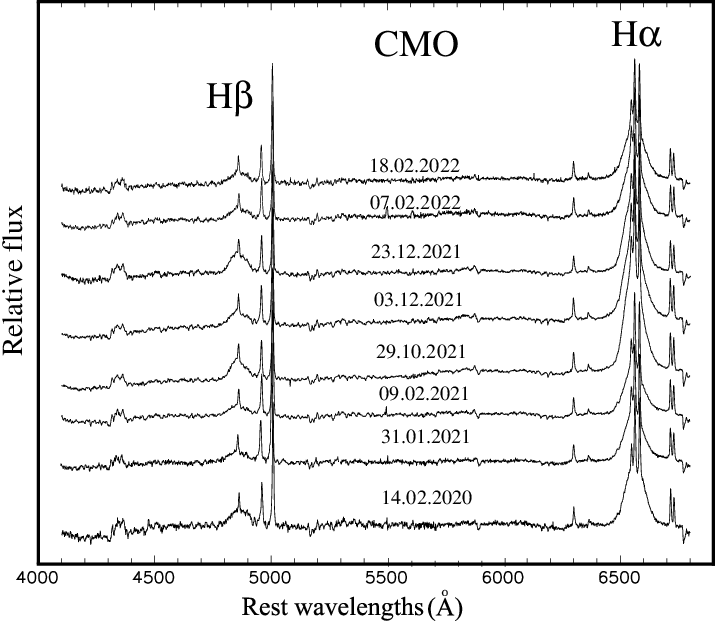}
 \caption{The H${\beta}$ -- H${\alpha}$ spectral region in the selected spectra obtained at CMO. The spectra are normalized to the [O III]${\lambda}$5007 intensity and shifted vertically for comparison. }
 \label{fig4}
\end{figure*}

\begin{figure}
\includegraphics[width=\columnwidth]{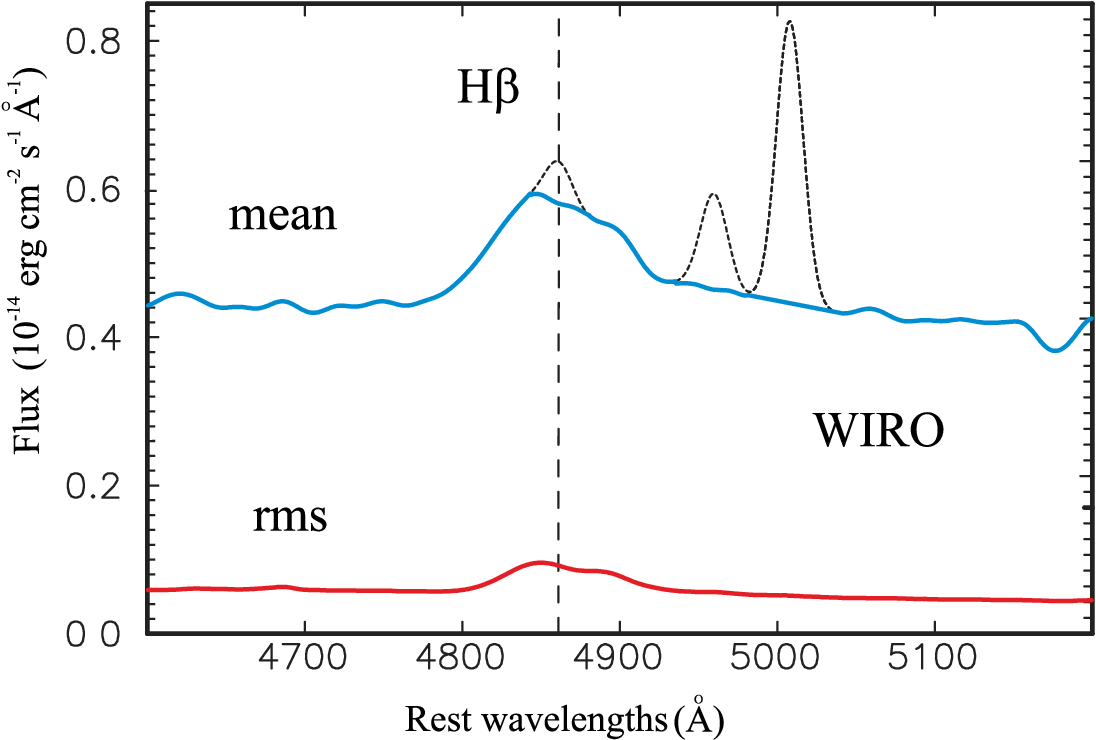}
  \caption{Mean (blue line) and rms (red line) spectra (WIRO) in the rest-frame for  the H$\beta$ region. In the dotted lines are the narrow H$\beta$ and [OIII]$\lambda$4959,5007 in the mean spectrum (see more details in the text),  while the dashed line indicates 4861~\AA ~in order to
illustrate the asymmetric H$\beta$ profile more clearly.}
   \label{fig5}
\end{figure}

\begin{figure} 
  \includegraphics[width=\columnwidth]{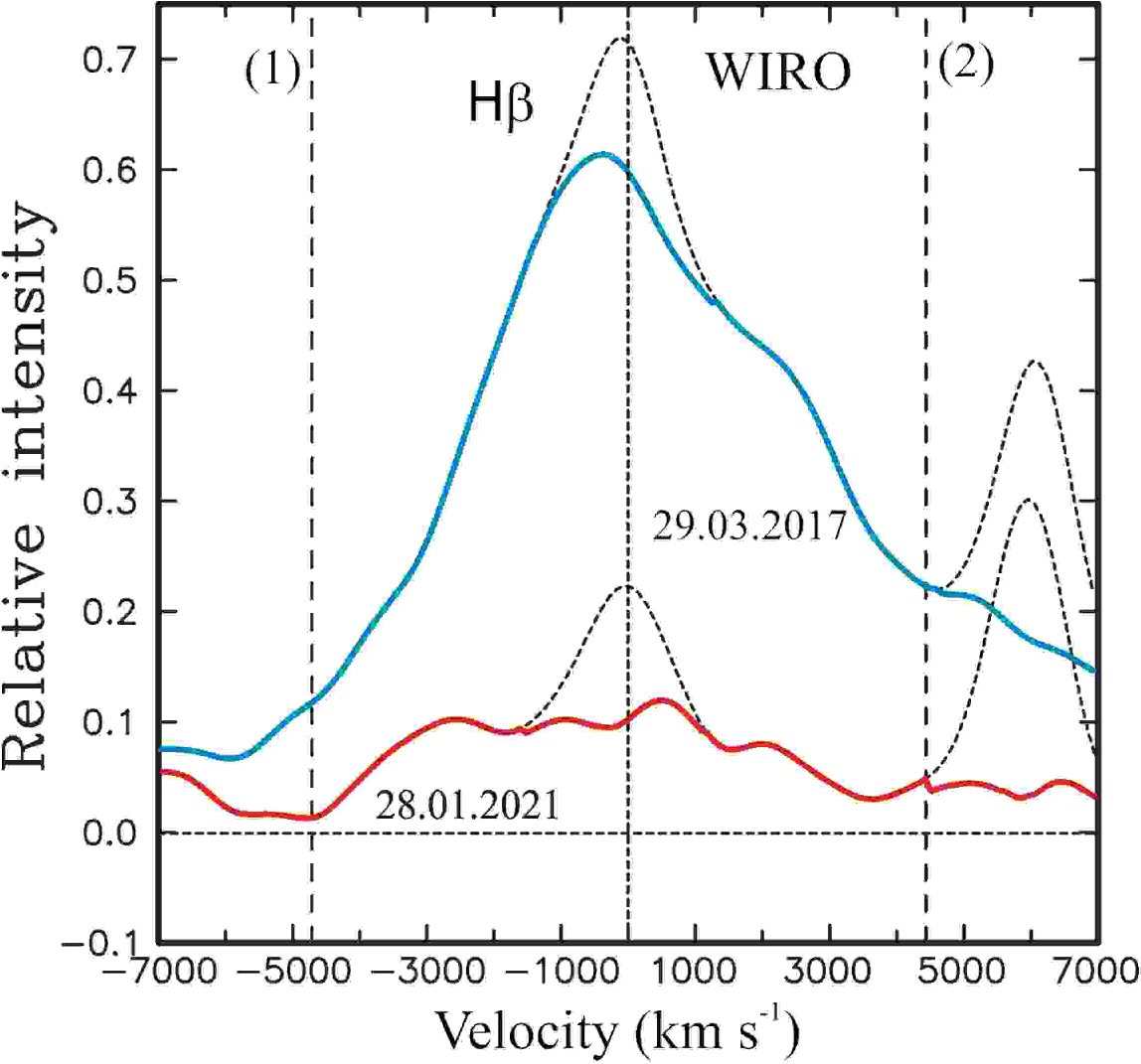}
    \caption{Profiles of broad H$\beta$ for the maximal (blue line)  and  minimal (red line) states with the narrow components removed showing intrinsic variations of the line (obtained at WIRO). 
    The vertical dashed lines (1--2) mark the range of integration. See text for details.}
    \label{fig6}
    \
\end{figure}

\begin{figure}
	\includegraphics[width=\columnwidth]{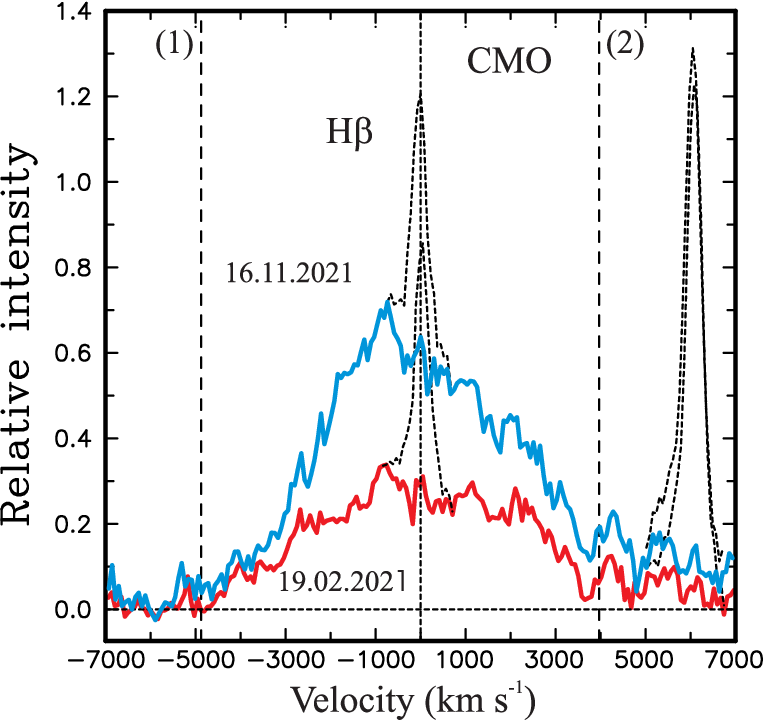}
    \caption{Profiles of broad H$\beta$ for the maximal (blue line) and  minimal (red line) states with the narrow components removed showing intrinsic variations of the line (obtained at CMO). 
    The vertical dashed lines (1--2) mark the range of integration. See text for details.}
    \label{fig5b}
    \
\end{figure}

\section{Broad emission-line profile variations}
During 2017-2022 a large variation in the H$\beta$ intensity (see Fig.~\ref{fig2}) was accompanied by a significant change in the line profile (see Fig.~\ref{fig3}  and Fig.~\ref{fig4}). The  mean and root-mean-square (rms)  spectra in H$\beta$ region (just WIRO data) for NGC~2617 are shown in Fig.\ref{fig5} (for details see \citealt{Du2018}). The  narrow emission lines in the rms spectra are not seen, since there is no variability and that is an indication of correct calibration procedure.  As seen from the rms spectra, the variability in the broad H$\beta$  line was significant and asymmetric. The examples of the H$\beta$ profiles for the maximum and minimum states are presented in Fig.~\ref{fig6} (WIRO) and Fig.~\ref{fig5b} (CMO). We remove the narrow component of the H${\beta}$ and the [OIII]${\lambda}$4959 line which blend with the broad component by subtracting copies of the [OIII]${\lambda}$5007 line, scaled and shifted to the best fit. A similar procedure was performed for H$\alpha$ in the CMO data, where the [OI]${\lambda}$6300 line was used as a template. For the WIRO data, the H${\alpha}$ flux is simply corrected by a constant contribution from the narrow lines, which is estimated using overlapping CMO and WIRO data. The profiles of broad H$\beta$ in the high state are more asymmetric with  the blue wing dominating. A double-peaked profile can be seen and the blue peak is closer to the narrow H$\beta$ component than the red one, and that is a reason why the red peak is seen more prominent in some spectra (see Fig.~\ref{fig3} and Fig.~\ref{fig4}).

\subsection{Variations of the H$\alpha$/H$\beta$ ratio}

\begin{figure}
	\includegraphics[width=\columnwidth]{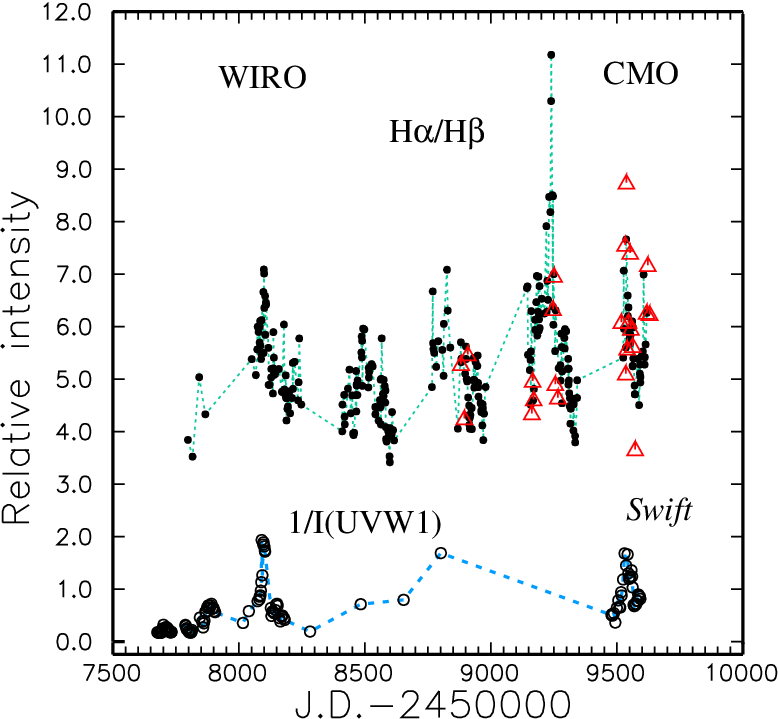}
    \caption{Variability of the broad H$\alpha$/H$\beta$ ratio from the WIRO (filled circles connected by green dashed line) and CMO (red triangles) spectra. For comparison, we plot the variations of 1/(I(UVW1)) from the $Swift$ data (open circles connected by the blue dashed line). See text for details.} 
   \label{fig11}
\end{figure}

The variations of the broad H${\alpha}$/H${\beta}$ ratio (based on the spectral data obtained at WIRO and CMO) are presented in Fig.~\ref{fig11}. The data were reduced to one system. The ratio variations are expected to anti-correlate with the UV flux level variability \citep[see, e.g.,][]{Ilich2012, Koll2018}. So, the variations of 1/(I(UVW1)--$I_h$(UVW1)) are shown in Fig.~\ref{fig11} for comparison, where $I_h$(UVW1) is the correction for the host galaxy contamination. The fast variations of both values are clearly visible on the timescale $\sim$ weeks in each season but some uncorrelated shifts appear if we consider all the data for a long time-scale ($\sim$year). The correlation coefficient is about 0.9 for each season but is much lower for all the data as a whole. The H${\alpha}$/H${\beta}$ ratio reaches the values about 8--11  in the 2020--2021 season. Such steep Balmer decrement values
are comparable with independent past results for other CL AGNs \cite[e.g][]{Osterbrock1981, Shapovalova2004, Koll2018}, however the errors in broad H${\beta}$  can be as large as $\sim$30\% or even more when the line is very weak.   
The steep Balmer decrement, namely the high H$\alpha$/H$\beta$ ratio in the minima of CL AGNs, can be attributed to a combination of low optical depth and low ionisation parameter in the BLR  as discussed for the cases of Mrk~1018, Mrk~609, Mrk~883, UGC~7064, Mrk~590 and NGC~1566 \citep{Osterbrock1981,Goodrich1995, Rudy1988, Denney2014, Oknyansky2019}.  The anti-correlation of the Balmer decrement and the continuum flux is a typical property of a CL AGN \citep{Shapovalova2004,Koll2018,Oknyansky2022}. Such property was found also for quasars more generally \citep{Ma2023}, however the amplitude of the effect is lower than for CL objects.  Variable obscuration is an alternative scenario and is often involved to interpret this type of variability \citep[see, e.g.,][]{Gaskell17}, but there are other possibilities \citep[see, e.g.,][]{Ilich2012, Korista2004}, which predict a steepening (flattening) of the broad-line Balmer decrement in low (high) continuum states.
This mechanism may explain fast variations (on a time scale of months) of Balmer decrement but also the two mechanisms  can act simultaneously and variable obscuration due to the sublimation and recovering of dust can explain the differences between seasons. 

\subsection{Implications for the black hole mass estimate}
The time delay of about $\tau$ $\sim$ 7.5 days (taking into account the $\sim$ 1.5 days delay for the optical and UV relative to X-ray variations) found for H$\beta$  ($R_{BLR}=c\tau$ is the emissivity-weighted radius of the BLR) and the full-width-half-maximum (FWHM) of the line rms-profile of about 4600 $km s^{-1}$ can be used to estimate the virial black hole (BH) mass:
\begin{equation}
\label{eqn1}
M_{BH} = f\frac{R_{BLR}\Delta V^2}{G}.
\end{equation}
where $G$ is the gravitational constant and $f$ is the virial factor determined by the geometry and kinematics of the BLR.
Using the same parameter $f=1.12$ as we used before \citep[for details and references see][] {Oknyansky2021}
we obtain a mass of  $\sim(3.5\pm1)\times10^7 M_\odot$ for the central BH. This value is generally consistent with the similar virial H$\beta$-based RM estimate of $(2.1\pm0.4)\times10^7 M_\odot$ published by \cite{Feng2021} and of $(4\pm1)\times10^7 M_\odot$ published by \cite{Shappee14}.
The relatively high precision of the above values reflects only measurement errors, while the systematic errors due to issues related to the choice of measurement algorithm, calibration of the $f$ factor and its deviation from the average \citep[e.g.,][]{Pancoast2014b}, as well as the choice of method for measuring the velocity $\Delta V$ \citep[e.g.,][]{Bonta2020}, make the masses much more uncertain.

\section{Changes of the UV/X-ray dependence and evolution of the X-ray spectrum}

\begin{figure}
	\includegraphics[width=\columnwidth]{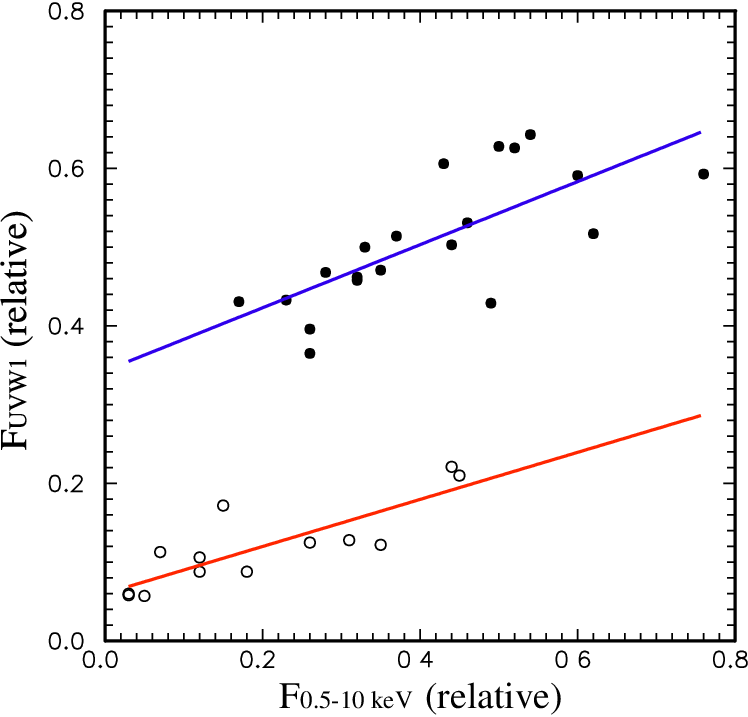}
    \caption{The F($UVW1$) versus F(X-ray) relation for two observational intervals: June 2016 -- April 2017 at the top (filled circles) and April 2017 -- January 2022 at the bottom (open circles). The straight lines show regressions obtained by the least square method.}  
   \label{fig13}
\end{figure}

In Fig.\ref{fig1}, we see a good correlation of fast variability in X-rays and  $UVW1$, but in recent years (2017--2022) there has been a slow downward trend in UV that has been absent in X-rays. Due to this fact, there has to be some evolution of the UV/X-ray flux ratio. To reveal this evolution, we took into account the existence of time delay between X-ray and UV variations of about 2 days and selected only those X-ray and UVW1 data that satisfy this delay.  In 
Fig. \ref{fig13}, the F($UVW1$) to F(X-ray) relations for two observational intervals (June 2016 -- April 2017  and April 2017--January 2022) are significantly  different, with a much lower UV/X(-ray) flux ratio for the second interval.

To trace the evolution of the X-ray spectrum of NGC~2617 as a function of luminosity, we performed an analysis following the procedure described by \cite{Oknyansky2021}. Namely, we selected two {\it Swift} data sets representing the states with minimal and maximal observed flux. For the low state, we took observations performed between December 4 and 17, 2017 (ObsID 00032812241--00032812246), whereas for the bright state we used the observation 00032812008 done on May 8, 2013. The fluxes in the 0.5--10 keV range were different in these states by a factor of 20. The following spectral analysis has been done using {\sc xspec} package \citep{1996ASPC..101...17A} and applying W-statistics \citep{1979ApJ...230..274W} after the energy spectra were rebinned to have at least one count per channel.

\begin{figure}
	\includegraphics[width=\columnwidth]{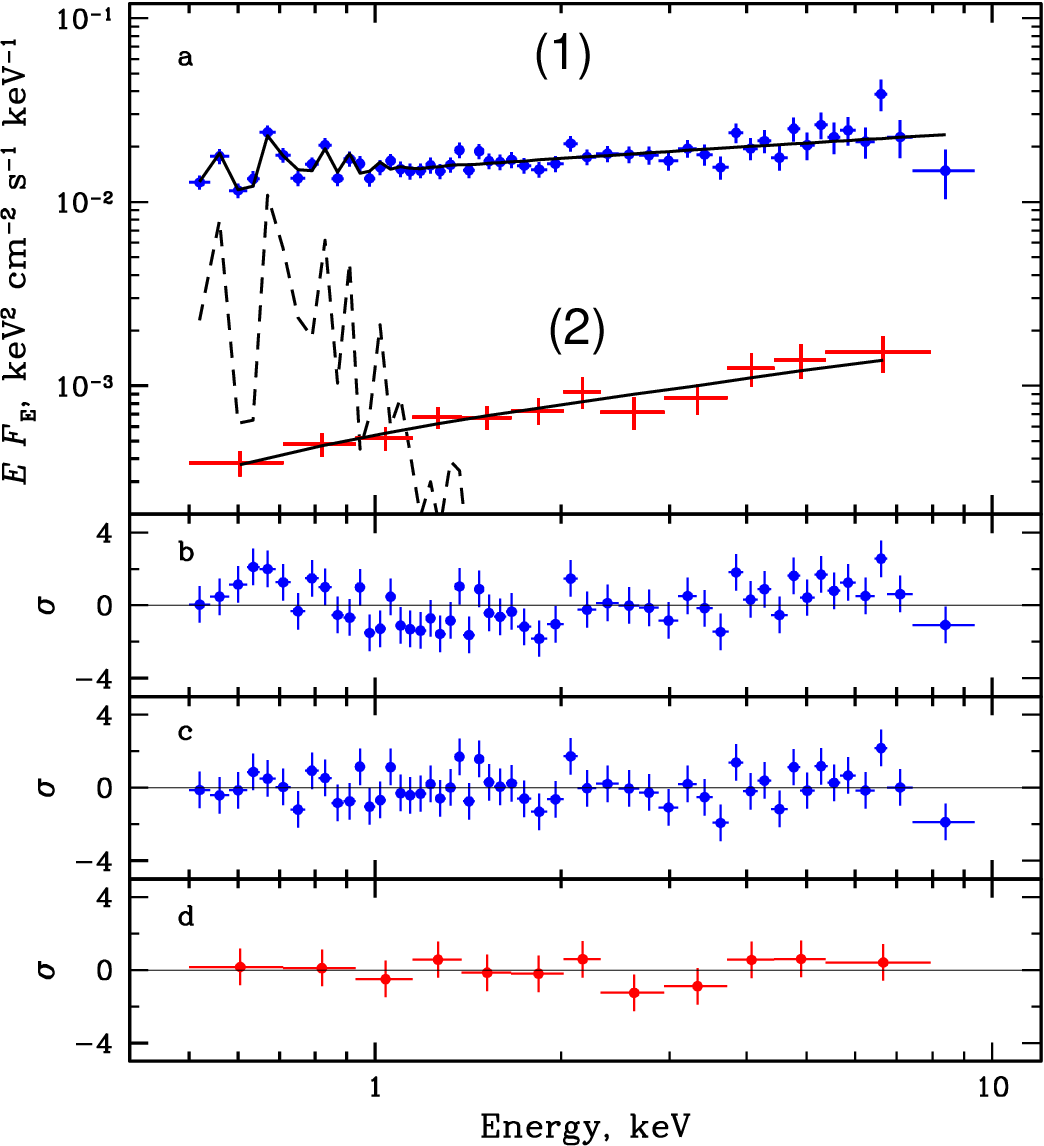}	
    \caption{(a) Unfolded X-ray spectra of NGC~2617 obtained with the {\it Swift}/XRT in the bright (ObsID 00032812008; (1) - blue points) and low (ObsID 00032812241--00032812246; (2) - red points) states. The best-fit models are shown with the solid lines. The additional soft emission component in the bright state is shown with the dashed line. (b) Deviations of data from the simple absorbed power law in the bright state. (c) Deviations of data from the absorbed power law with soft emission component in the bright state. (d)~Deviations of data from the simple absorbed power law in the low state.
    }  
   \label{fig14}
\end{figure}

For the first step we fitted spectra in both states with the simplest absorbed power-law model ({\sc phabs$\times$po} in {\sc xspec}) with the absorption fixed at the Galactic value of 3.7$\times10^{20}$ cm$^{-2}$ \citep{2016A&A...594A.116H}. The unfolded spectra are shown in Fig.~\ref{fig14}(a) with the best-fit parameters of the simple model presented in Tab.~\ref{tab:spe}. As can be seen from panel (b) of the same figure, some evidence for a soft emission component can be noticed around 0.6--0.7 keV in the bright state. To improve the quality of fitting we introduced a {\sc mekal} component with a temperature of $0.25\pm0.02$~keV and normalization $5.2\pm1.1$. This modification decreased the C-statistics value to 505.7 with 569 d.o.f. and made photon index slightly harder ($1.80\pm0.4$). Unfortunately, the limited energy range and low count statistics do not allow us to make final conclusions about the composition of the intrinsic spectrum of the source even in the bright state. However, the hardening of the spectrum towards the low state is detected reliably. Similar behaviour was found for several CL AGNs and can be caused by different reasons \citep[see, e.g.,][for a review]{Ricci2022}.

We can conclude from the available data that the soft thermal component in the bright state can not be explained by the contribution of the host galaxy. The  significant difference detected in the photon index values  (for high and low states) can be explained, e.g., by a substantial drop of the accretion rate in the low state and following change of the ionisation state in the material nearest to the SMBH (supermassive black hole), the Compton-thick clouds  which partially obscure the X-ray source. In high states these Compton-thick clumps are highly ionised and in case of crossing the line-of-sight may produce temporarily colorless obscuring of the X-ray source \citep[see, e.g. Paper I;][]{Oknyansky2021, Turner2011}. In low states these clumps are less ionised and may produce obscuration with change of the photon index. 

\begin{table}
  \noindent
\caption{Best-fit parameters for a {\sc phabs$\times$po} model obtained for two {\it Swift}/XRT observations. All errors are reported at 1$\sigma$ confidence level.}
\label{tab:spe}
\begin{tabular}{ccc}
\hline
\hline
Parameter &     Units                   &           Value \\
\hline
\multicolumn{3}{c}{High state (ObsID 00032812008)}  \\
\hline    
$N_{\rm H}$ & $10^{20}\text{ cm}^{-2}$ &   $3.7^{fixed}$ \\
Power-law ph. index     &               &   $1.94\pm0.02$   \\
Power-law norm.      &                  &   $(1.74\pm0.03)\times 10{-2}$   \\
Flux$^a$ &  $10^{-11}$ erg s$^{-1}$ cm$^{-2}$ & $8.4\pm0.1$\\
C-statistics (d.o.f.) &  & 530.1 (573)\\
\hline

\multicolumn{3}{c}{Low state (ObsID 00032812241--00032812246  )}  \\
\hline 
$N_{\rm H}$ & $10^{20}\text{ cm}^{-2}$ &  $3.7^{fixed}$ \\
Power-law ph. index     &               &   $1.54\pm0.08$   \\
Power-law norm.      &                  &   $(5.8\pm0.4)\times10^{-4} $   \\
Flux$^a$ & $10^{-12}$ erg s$^{-1}$ cm$^{-2}$ & $4.3\pm0.3$\\
C-statistics (d.o.f.) &  & 198.1 (240)\\

\hline
\end{tabular}
	\begin{tablenotes}
        \item $^{a}$ Observed flux in the 0.5--10 keV energy band.
    \end{tablenotes}
\end{table}

\section{Discussion}

NGC~2617 is one of the clearest cases where the appearances of both type 1 and type 2 AGNs have manifested themselves in the same object at different epochs. What must happen to make such a dramatic change possible? CL AGNs like NGC~2617 present a problem for the simplest unification model that considers the type 1 and type 2 AGNs to have the same nature and the difference in type to be due to the orientation to the observer. In this picture, we see a type 2 AGN if the BLR and AD are blocked from our view by obscuring dust surrounding the AGN perpendicular to the axis of symmetry \citep{Keel80}. However, the orientation cannot change on the timescale of the observed type changes and hence some other explanation is needed.  In a review by \cite{Ricci2022}, CL AGNs were divided in 2 classes: (1) changing-obscuration objects and (2) and changing-state AGNs. We have suggested \citep{Oknyansky15} that CL AGNs can have a combination of these two factors: i.e., a CL event can be due to a significant change in the accretion rate which leads to a variable absorption of the continuum and broad line emission by dust. We proposed a hollow bi-conical dust distribution model \citep{Oknyansky15,Oknyansky2018b, Oknyansky2023}), where the sublimation or recovering of dust in some clouds along the line of sight in the hollow cone can be an additional factor to explain the CL phenomenon \citep[see also investigations of  pole dust obsucration in AGNs,][]{Gaskell2018,Buat2021}.  The model can explain not only the similar IR lags
at different wavelengths \citep[see results of numerical  simulations performed by][]{Oknyansky15} but also the high correlation values for the IR and optical (UV) fluxes. Furthermore, the model can better (than in case of flat geometry) explain the observed luminosity in the IR due to a higher covering factor for UV radiation. The need for fast re-formation of the dust in order to explain the good correlation between UV and near-IR light curves was realized three decades ago \citep{Barvainis1992} and later examples of the dust recovery were observed for some CL AGNs \citep{Oknyansky1999, Oknyansky2006, Oknyansky2008, Oknyansky2014a, Kokubo2020}. There are two main mechanisms that may be responsible for the recovery of dust after sublimation: (1) a condensation mechanism similar to the formation of dust in novae and supernovae; (2) delivery of dust from farther distances from the central source. Besides, it is possible that some amount of the dust survives in dense clouds during the outbursts being shielded from the UV radiation and then somehow moves from there to the outer parts of the cloud \citep{Barvainis1992}. In any case, this process takes some time. The estimates of the time required for the dust to recover were based on observations, and RM for different time intervals gave a value on the order of several years  \citep{Oknyansky2006, Oknyansky2008, Oknyansky2014a, Kokubo2020}. Our results demonstrating the decrease of the near-IR time delay for NGC~2617, which started after a few years of being in the low state, are in agreement with the proposed recovery of the dust. In the conical outflow model, depending on the recent luminosity history, the opening angle of the hollow, dust-free conical region (as well as the same for the partially dust sublimated conical region) will be larger or smaller \citep[see Fig. 1  in ][]{Oknyansky2023} due to dust recovery or sublimation processes. That can explain saturation in the relation between $F{_\lambda}(B)$ 
 (which is proportional to $F_{UV}$) and  $F{_\lambda}(K)$ at Fig.~\Ref{ibk}, which can be connected with variations of the covering factor.
In Paper I, using the relation from \cite{Sitko1993}  (see  Eq. 1 and details there), we obtain a rough estimate for the dust sublimation radius (R$_{sub}$) at NGC2617 of about 32 light-days.
Due to several not well known parameters, this estimate only provides a very preliminary value, but it is close to the time delay for $K$  band \citep[this relation with the same parameters well fits the observaional data,][]{Oknyansky2001}.
In the conical dust outflow model, the sublimation radius R$_{sub}$ also depends on the effective half-opening angle  of the dust conical region   due to axial symmetry of the UV radiation \citep[see, e.g.,][]{Netzer2015} as well as due to absorption of the UV radiation by dust.  
The change in the mean UV luminosity (during 2016-2018 and 2019-2021)  was very significant (factor of $\sim~4$)  and that can explain variations of about a factor of two in the time delay, but we have to consider possibilities for fast dust recovery \citep[see, e.g.,][]{Barvainis1992, Oknyansky2023}.
More detailed discussion of the conical dust outflow model will be done in future publications.

Changes in energy generation in CL AGNs are common, although it is not clear what drives them \citep[see discussion and references on the topic in][]{MacLeod2019, Oknyansky17, Oknyansky2019, Oknyansky2021, Runnoe2016, Ruan2019, Hon2022}.  Mostly the discussed mechanisms responsible for such dramatic changes are various types of instabilities in the AD and tidal disruption events (TDEs). The TDE explanation has a problem of too low predicted cadence of such events that contradicts the fact that the observed CL phenomena are not too rare, as well as the recurrence of the events in individual objects \citep{Oknyansky2022}. A similar mechanism involving the tidal stripping of stars \citep{Campana15, iv06} could lead to events more frequent (than TDEs) and recurrent events of significant brightening. The recurrence of CL events can find a natural explanation in models with AD instabilities \citep [see, e.g.,][]{Sniegowska2020} and also by invoking the case of close binary black holes with low mass ratio \cite{Wang2020}. CL AGNs were proposed by \citet{Wang2022} to be some particular AGNs at a special evolutionary stage, such as a transition stage from different levels of fueling of the supermassive black hole. \citet{Dodd2021} investigated the properties of the host galaxies of CL AGNs and found suggested certain physical conditions leading to episodic accretion events responsible for triggering CL activity. 

For NGC~2617, the duration of monitoring is too short to be able to fix the recurrence of CL events. It is unclear if the object had been in the low state characterized by spectral type Sy1.8 permanently or for a long time before this interval of high activity started, but we do see some evidence that the object is moving back to the low state. We have detected very deep minima in all wavelengths as well as in broad emission lines.   It is useful to continue the monitoring which can help in selecting the best model to explain the CL phenomenon in this case.  

We have found time delays between continuum variations at different wavelengths (X-ray, UV, optical, near-IR), as well as delays between continuum and broad emission lines, using our own statistical method MCCF. It is common to attribute the optical/UV emission to viscous heating of matter in an optically thick AD around a SMBH \citep{1_sunyaev73, Novikov1973}. The commonly invoked `lamppost' model predicts an X-ray emitting `corona' located on the rotation axis of the SMBH \citep[see, e.g.,][]{Cackett2007} and wavelength-dependent time delays between the continuum at different wavelengths due to reprocessing of the extreme UV/X-ray photons into UV/optical radiation arising in the AD. A delay of $\sim$ 2 days found in this study for the lag of UV variations relative to the X-rays (which is in agreement with previously published results \cite{Shappee14, Kammoun2021}), in combination with a very short delay between the UV and optical variations, is in contradiction with the lamppost-type X-ray reprocessing model. The same problem was noted before as a common property of CL AGNs \citep[see, e.g.,][]{Cackett2007,  bu17,  Edelson2017, McHardy2014, Shappee14, Oknyansky17, Oknyansky2020c, Oknyansky2021, Oknyansky2022,Lawther2022}. The sizes of ADs measured from the microlensing data for some gravitationally lensed quasars are also significantly larger than those predicted  by the reprocessing model \citep[e.g.,][]{Morgan2010}. A possible version of the lamppost-type scenario is to place the X-ray source at a large height (about 2 light days) from the SMBH, however this option was discussed by \cite{nd16} and rejected as highly unrealistic (see details therein). Two alternative explanations for the obtained results are: (1) a truncated AD in analogy with black-hole binaries (e.g, \citet{Done2007}) and (2) the model with a soft excess region at the inner edge of an AD which completely hides the hard X-ray corona \citep{Gardner2017, Edelson2017}. Both possibilities were discussed by \cite{nd16} and they preferred the first explanation (1).

Meanwhile, our results for NGC~2617 (as well as for NGC~3516) involving soft variable black body components contradict the truncated AD model and are in agreement with the predictions of the second explanation (2) (see details in \citealt{Oknyansky2021}). Using our optical spectroscopy and multi-wavelength photometry (from near-IR to X-ray region), we have shown that NGC~2617 had been in the high state for several years (2013--2017) (see also Paper~I), and from the beginning of April 2017 until the end of May 2018 the object had very low brightness and variability. In December 2017, the X-ray flux was the lowest since monitoring began.  Despite the very low X-ray and UV flux observed in December 2017, the broad H${\beta}$ line was still prominent. As it was suspected in our publication (Paper I), the dust needs at least a few years of the object being in the low state to recover and to be able to obscure broad lines. Following this prediction the object was for a few years relatively low (2019--2021) before a new CL event was detected for the object (a change from Sy1 to Sy1.8 spectral type  and this CL transition can
be numerically indicated as +2,  in the “magnitude” system introduced by \cite{Runco2016}). We have also found that the near-IR time delay became significantly lower than it was in 2016. Variable dust obscuration may explain the decrease of the UV/X-ray flux ratio in 2017--2022 found in our study. The tendency for a higher H${\alpha}$/H${\beta}$ ratio may also exist due to variable dust obscuration. All these facts support the assumption of dust recovering during the long low state of the object.

\section{Summary}

Using spectroscopy and multi-wavelength photometry, we have shown that NGC~2617 continued to be in the high state in recent years and to appear as a type 1 AGN, but more recently (2019--2021) the object was relatively low and the broad H$\beta$ line has become very weak.  Perhaps as we proposed in Paper I, the object needs to be in the low state at least for a few years so that the dust can recover and obscure broad lines. The object has fulfilled the prediction and a new CL event was detected. We have also found that near-IR time delay became significantly lower than it was in 2016, which also supports the assumption of dust recovering. In the case of absorption, the UV/X-ray ratio also has to drop down. The duration of the high state and the continuing variability are not readily consistent with the type change being due to a tidal disruption event or a supernova. We propose that the original change of type was the result of increased luminosity causing the sublimation of dust in the inner part of a bi-conical dusty outflow. This leads to a much clearer visibility of the central regions. The recovery of dust occurring during the long low state is a reason for a new CL event.

\section*{Acknowledgements}
 We thank  A.~Cherepashchuk for supporting our research and observations, and the staffs of the observatories. We also express our thanks to the {\it Swift} administrators for approving the ToO observation requests and the {\it Swift} ToO team for promptly scheduling and executing our observations. We are grateful to M.~Gaskell, P.~Ivanov, H.~Netzer and A.~Laor for useful discussions. This work and MASTER equipment were supported in part by the M.~V. Lomonosov Moscow State University Program of Development, by the National Research Foundation of South Africa. This research has been partly supported by Israeli Science Foundation grant no. 2398/19 and by the Center for Integration in Science of the Ministry of Aliyah and Integration.  We thank WIRO engineers James Weger, Conrad Vogel, and Andrew Hudson for their indispensable and invaluable assistance. M.S.B. enjoyed support from the Chinese Academy of Sciences Presidents International Fellowship Initiative, grant No. 2018VMA0005. T.E. Zastrocky acknowledges support from NSF grant 1005444I.

\section*{Data availability}
The data underlying this study are available in the main body of the article and in online supplementary material.




\bibliographystyle{mnras}
\expandafter\ifx\csname natexlab\endcsname\relax\def\natexlab#1{#1}\fi
\bibliography{NGC2617-2021} 

\bsp	
\label{lastpage}
\end{document}